**Digital Nature Revisited: A Ten-Year Synthesis of Art, Technology, and the Evolution of 'Nature'**

Reimagining Post-Truth Ecologies Through Art, Algorithm, and Animism

Yoichi Ochiai[*1] and Takashi Shimizu[*2]

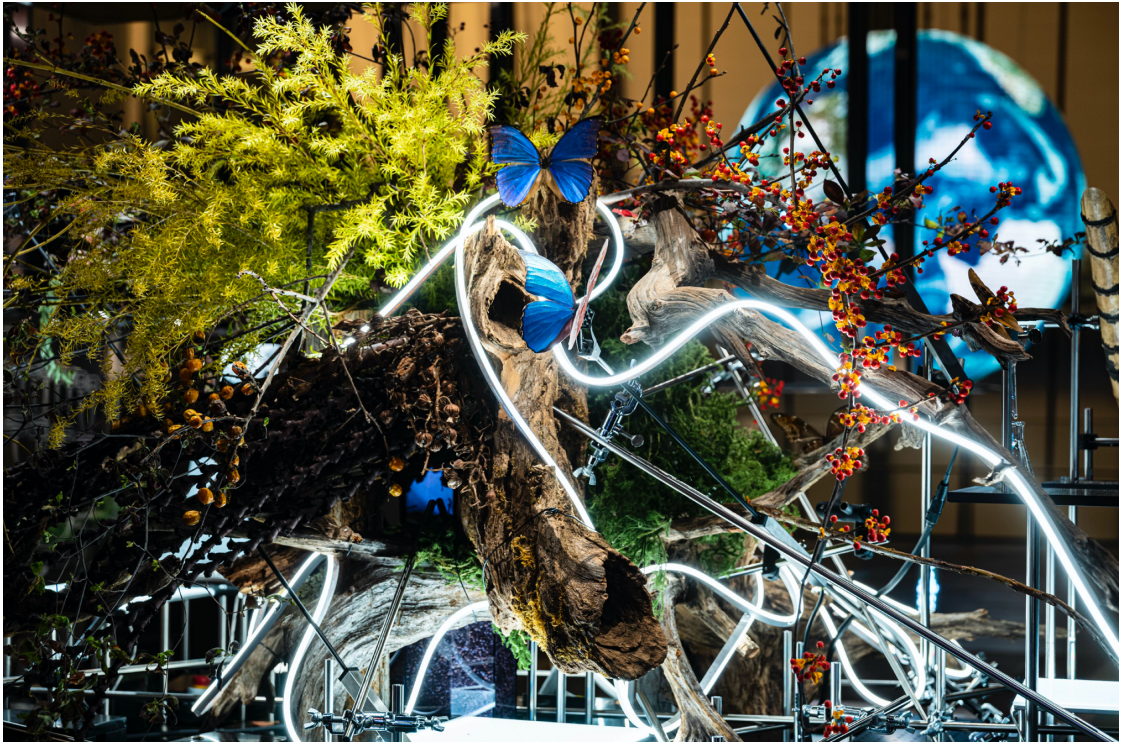

Figure 1: This scene captures the coexistence of a natural Morpho butterfly, an artificial structural color Morpho butterfly created through lithography, natural flowers and robots, and an LED Earth.  The Photo taken in the exhibition of Digitally Natural, Naturally Digital at Miraikan - The National Museum of Emerging Science and Innovation, Japan.

This paper critically re-examines "Digital Nature," a concept that has proliferated across various domains over the last ten years. By "Digital Nature," we refer to an evolving view of nature as a dynamic process of circulating computation and matter—one that extends into the realms of AI, XR, indigenous perspectives, and post-human theory. Despite its popularity, "Digital Nature" remains ambiguously defined. This paper provides a genealogical and philosophical survey of how the idea has emerged, diverged, and overlapped in media art, bio-art, and generative art, alongside relevant Eastern, Islamic, and indigenous worldviews. We then introduce a multi-axis framework (from real/virtual to anthropocentric/object-oriented, with sub-axes of enchantment and materialization), illustrating how digital technologies have reconceptualized the question "What is nature?" in unexpected ways. Finally, we discuss how the field might evolve, particularly through the lens of large language models, AGI, and "supernatural reality," while highlighting the ethical and political pitfalls of techno-occultism. Our ultimate goal is to re-situate "Digital Nature" as both an intellectual frontier and a collaborative platform that invites continuous dialogue between art, science, technology, and cultural philosophies.

CCS CONCEPTS • Applied computing  • Arts and humanities • Media arts

**Additional Keywords and Phrases:** Digital Nature, Object-Oriented Ontology, Post-humancentrism

*1: University of Tsukuba, wizard@slis.tsukuba.ac.jp
*2: Toyo University

# 1 INTRODUCTION

The term "Digital Nature" has been used in many overlapping ways over the past decade. It frequently appears in contexts such as media art, bio-art, AI-generated art, and computational design, often signifying a "fusion of nature and the digital." Yet behind this seemingly straightforward phrase lies an array of philosophical, cultural, and scientific complexities. While some projects employ "nature" primarily as a visual spectacle (e.g., mesmerizing fractals in projection-mapped art installations), others emphasize a deep reflection on what "nature" actually means in an age when computation is woven through every dimension of ecological and social life.

From its early roots in computer graphics (CG) research, where fractal terrain generation and Perlin Noise shaped virtual landscapes, to more recent explorations of posthumanism, indigenous cosmologies, and generative adversarial networks (GANs), the discourse around "nature" is now evolving into an interdisciplinary junction that merges design, engineering, biology, philosophy, and spirituality. As artificial intelligence (AI) tools become more advanced, the boundaries between "natural" and "artificial" have grown increasingly porous. Meanwhile, social and ecological challenges—from climate change to the commodification of living systems—necessitate robust critical debate on how nature is simulated, augmented, or re-enchanted through digital means.

In this paper, we offer an integrative perspective on "Digital Nature" by (1) mapping its key genealogies and philosophical influences, (2) introducing a framework for analyzing current works and research, (3) performing cross-comparisons of exemplary projects, and (4) considering the implications of future technologies such as large language models (LLMs) and potential AGI. By unearthing both the historical and cultural contexts, we hope to clarify how different communities, from media artists to environmental scientists, are co-producing new conceptualizations of nature that oscillate between enchantment and implementation, politics and aesthetics, tradition and innovation. We aim to show that the ambiguity of "Digital Nature" is not merely an obstacle but can serve as a generative force for exploring how matter, computation, and living systems interrelate.

## 1.1 Research Questions and Contributions

- RQ1: Historical and Conceptual Scope. How has the notion of "Digital Nature" emerged from early computer graphics research and generative art, and why has it expanded across diverse fields (media art, bio-art, VR/AR, environmental sensing)?
- RQ2: Framework Development. Can we introduce a conceptual scheme that captures the range of "Digital Nature" practices—from anthropocentric to object-oriented perspectives, and from simulation to materialization—while addressing spirituality, mysticism, or socio-political structures?
- RQ3: Toward a Pluralistic Ecology. What do non-Western philosophies (Eastern, Islamic, and indigenous worldviews) contribute to this redefinition of nature, and how do these converge with computational or generative paradigms?
- RQ4: Critique and Future Possibilities. How does the rising wave of AI, particularly LLMs and other advanced models, reshape the prospects of "Digital Nature," and what are the cultural or ethical risks of "techno-occultism"?

By exploring these questions, the paper integrates artistic, technical, and philosophical debates. We propose that "Digital Nature" is not just a convenient catchphrase but a critical space to re-examine the interplay of nature, computation, and culture in the era of advanced digital technologies.

## 2 GENEALOGY AND CURRENT STATUS OF DIGITAL NATURE

### 2.1 Early Explorations in Computer Graphics

The early background of "Digital Nature" can be traced to the development of computer graphics techniques designed to simulate natural phenomena. Ken Perlin's noise function [1], for instance, revolutionized texture synthesis, allowing more organic visuals in movies and video games. Around the same time, Benoit Mandelbrot's fractal geometry of nature [2] greatly influenced artists and researchers seeking algorithmic representations of mountains, coastlines, and clouds. The SIGGRAPH community in the 1980s and 1990s embraced such explorations, broadening the conversation to include generative aesthetics and the "algorithmic beauty" of natural forms [3].

Beyond mere visual realism, some pioneers recognized deeper conceptual implications: if nature could be described algorithmically, perhaps "nature itself" was, in part, computational. This led to natural computation and artificial life research, in which fractals, L-systems, and agent-based simulations served not merely as visual tools but as ways to explore "organic" or emergent behaviors in a digital substrate [4–6].

### 2.2 Generative and Bio-Art

By the 2000s, generative art became well established, bringing evolutionary algorithms, randomness, and complex systems into artistic practice [7,8]. Simultaneously, bio-artists began working with living materials (e.g., bacteria, cells, plants) as media, framing "nature" as an active collaborator rather than a passive resource [9,10]. These developments continued to blur lines between natural and artificial, living and computational, thereby laying fertile ground for the concept of Digital Nature to emerge more explicitly.

In addition, with the proliferation of Augmented Reality (AR) and Virtual Reality (VR), new ways of "experiencing nature" in virtual space proliferated [11,12]. Some researchers explored "Digital Urban Nature," investigating how apps, sensors, and big data could integrate or replicate nature in urban contexts to foster well-being or ecological awareness [13,14]. Critiques arose, warning that such initiatives might reduce nature to a commodity for human-centered use, even if couched in ecological rhetoric [15].

### 2.3 AI, Large Language Models, and Posthuman Perspectives

The 2010s saw deep learning, particularly generative adversarial networks (GANs) [16], dramatically expand capabilities to synthesize or manipulate images of nature. Meanwhile, large language models (LLMs) advanced the capacity to generate textual or multimodal representations of landscapes, ecosystems, or entire artificial worlds [17]. Posthuman theorists [18,19] and indigenous scholars [20] argued for broader definitions of nature that include non-human agencies and multi-species entanglements. Artists like Yoichi Ochiai [21,22] and Refik Anadol explored visions of "Digital Nature" blending data visualizations, immersive installation, and references to spiritual or mythic dimensions.

Yet, the term often remains ambiguous. Some see it as simply a new marketing-friendly label for VR or interactive art installations featuring plants and waterfalls. Others view it as a radical turn in how we conceptualize ecological systems, mysticism, embodiment, or animistic cosmologies through computational means [23].

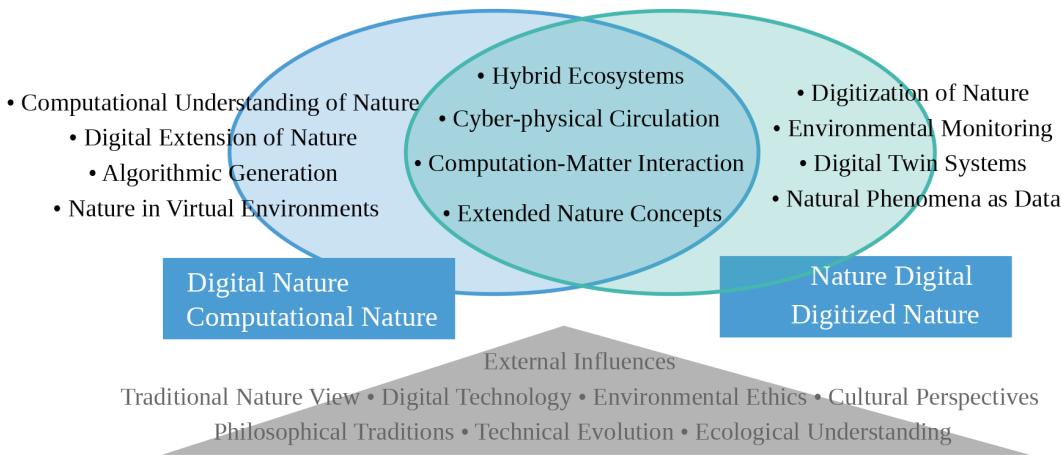

Figure 2: Background of Digital Nature's Genealogy - A Summary of Thee Major Trends

## 3 HISTORICAL AND PHILOSOPHICAL FOUNDATIONS AND PLURALISTIC VIEWS OF NATURE

### 3.1 Serres, Leibniz, and Bachelard: Discontinuities and Emergence

Michel Serres's works on "discontinuity" and "recursive models" [24] provide a lens through which to re-interpret "Digital Nature." Rather than seeing knowledge development as linear, Serres argues that ruptures, digressions, and overlaps generate new insights. This resonates with the emergence of "Digital Nature," where prior frameworks of "nature vs. machine" give way to re-combinations and conceptual leaps. Serres was deeply influenced by Leibniz, whose monadology viewed the universe as a network of monads each reflecting the entire cosmos in its own perspective [25]. Applied to "Digital Nature," this monadological reading suggests that a piece of digitally simulated "nature" might reflect broader ecological or philosophical structures—even if it is only partial, discrete, or seemingly artificial. Gaston Bachelard's emphasis on epistemological breaks and the imaginative leaps necessary for scientific revolutions [26,27] similarly points to the idea that "nature" in scientific discourse is constantly being reconfigured. Bachelard argued that illusions and poetic intuitions can be drivers of scientific innovation, hinting that generative art or immersive VR might not merely be ephemeral spectacles but can also transform how we conceptualize ecology, matter, or even spirituality.

### 3.2 Pluralistic Approaches to Nature: Eastern, Islamic, and Indigenous Views

#### 3.2.1 Eastern (Taoist and Buddhist) Perspectives

Taoist philosophy, particularly the concept of wu wei (non-action), offers a worldview where humans and nature are not in a dominative relationship but in a reciprocal, spontaneous process [28]. In Buddhist thought, the principle of dependent origination indicates that every phenomenon arises in relation to others, suggesting a "network ontology" that is strikingly analogous to certain computational or ecological models. Rather than focusing on controlling or harnessing nature, these perspectives emphasize attunement and interdependent co-arising [29,30].

When applied to Digital Nature, these Eastern philosophies encourage designing computational systems that foreground relational processes, subtle emergences, and balanced minimalism, rather than imposing top-down control or turning nature purely into data. One can see these ideas in some interactive art installations that let natural or semi-random processes evolve with minimal human interference.

### 3.2.2 Islamic Geometry and Mystical Cosmology

In Islamic traditions, geometry has historically been used to symbolize divine order without explicitly depicting living forms [31,32]. Contemporary digital artists have leveraged algorithms to produce intricately repeating patterns akin to Islamic ornamental designs, pointing to a worldview where mathematical precision and spiritual depth intertwine. Some interpret these geometric structures as bridging the gap between a "computable cosmos" and the ineffable or mystical—a resonance that can undergird certain "supernatural" threads in Digital Nature [33].

### 3.2.3 Indigenous Thought and Animism: Inseparable Continuity with Nature

For many indigenous worldviews, all entities in nature—animals, plants, stones, rivers—carry some form of agency or spirit [20,34]. Humans are neither outside nor above nature but entangled with it in continuous relationality. In the digital realm, such animistic views can challenge anthropocentric assumptions by inviting designers and artists to create multi-agent simulations or interactive systems that do not privilege human control. Instead, non-human agents (virtual or biological) can become co-producers of aesthetic or ecological processes [35]. This has implications for how we approach "Digital Nature" ethically and politically, especially in the face of climate change and environmental degradation.

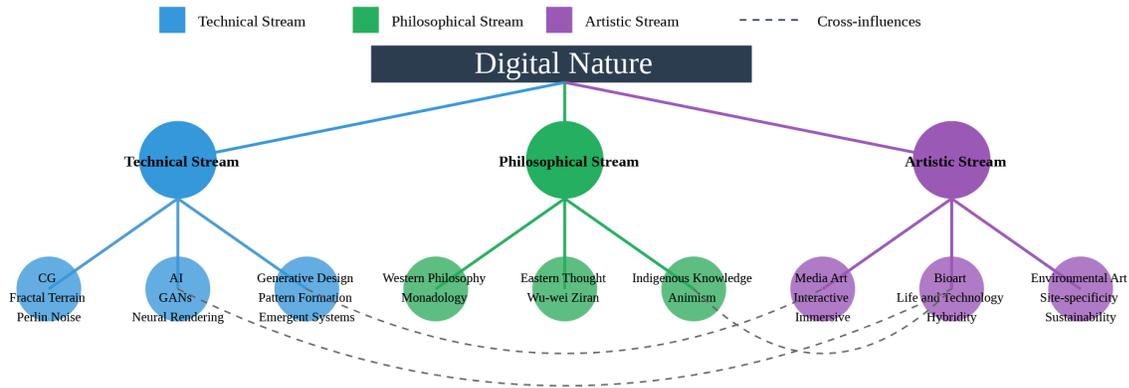

Figure 3: Evolution of Digital Nature Theory: Theoretical Development and Conceptual Evolution

## 4 CASE STUDIES AND MULTI-AXIS FRAMEWORK

### 4.1 Proposed Framework: Four Quadrants, Eight Subcategories

To address the broad array of Digital Nature works, we propose mapping them onto two major axes: Anthropocentric vs. Object-Oriented (i.e., who or what is centered as the primary agent/beneficiary) and Enchanted vs. Implemented (i.e., whether works emphasize spiritual, mystical, or aesthetic enchantment vs. practical, solution-oriented aspects). Each quadrant can then be subdivided by a distinction between Simulated (primarily algorithmic, virtual) vs. Materialized (involving physical, biological, or tangible elements).

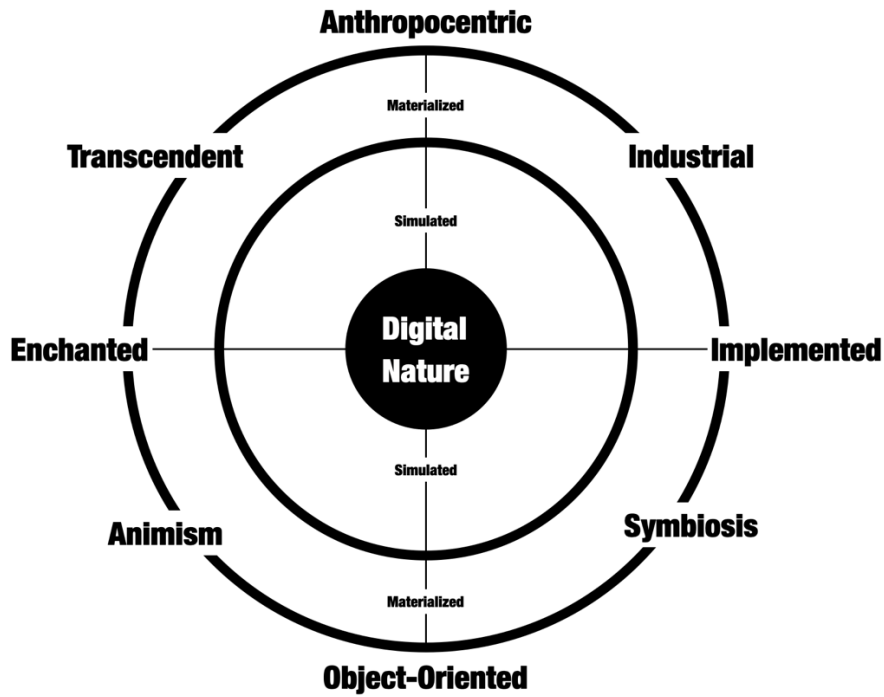

Figure 4: Anthropocentric–Object-Oriented × Enchanted–Implemented × (Simulated/Materialized)

1. **Q1: Transcendent (Anthropocentric + Enchanted)**
    - *Simulated*: VR installations where a user experiences a "sacred garden" or mythic realm, focusing on transcendent human feelings.
    - *Materialized*: AR pilgrimage apps layering digital mysticism onto real sacred sites, heightening human ritual experience.
2. **Q2: Industrial (Anthropocentric + Implemented)**
    - *Simulated*: Large-scale simulations for optimizing resource distribution, urban green spaces, or logistic networks for human utility.
    - *Materialized*: Smart farms using sensors/drones to mass-produce crops, focusing on efficiency and controlling natural processes for industrial gain.
3. **Q3: Symbiosis (Object-Oriented + Implemented)**
    - *Simulated*: Eco-simulations exploring multi-agent coexistence, where AI-driven animals, plants, and humans share an environment in near-equal footing.
    - *Materialized*: Urban biodiversity projects or sensor-driven community science initiatives that treat animals, plants, or microorganisms as co-creators.
4. **Q4: Animism (Object-Oriented + Enchanted)**
    - *Simulated*: Virtual worlds where multiple "non-human avatars" are endowed with distinct spiritualities or mythic qualities, inviting participants to experience "nature's soul."
    - *Materialized*: Bio-digital installations blending indigenous ritual or animistic worldviews with electronic or interactive systems. The physical environment itself is treated as spiritually alive, not a mere resource.

By situating projects within these eight subcategories, researchers can identify convergences, divergences, and possible blind spots. For instance, a "Digital Nature" project that claims to celebrate harmony with nature might in fact be Industrial + Implemented (Q2) if it predominantly harnesses nature for anthropocentric objectives. Conversely, an experience that looks purely decorative might have deeper ties to indigenous or animistic frameworks.

### 4.2 Examples of Projects

- Transcendent-Simulated: Some VR installations replicate religious or mythical landscapes, letting users undergo an interactive pilgrimage. While these often revolve around human emotional engagement, they can open contemplative or mystic dimensions.
- Industrial-Materialized: Vertical farms using IoT to regulate water/nutrients exemplify a deeply human-centered approach to controlling natural processes. While beneficial for food production, they rarely incorporate an animistic or spiritual perspective.
- Symbiosis-Simulated: Projects that simulate multiple species' habits or pollination patterns with AI agents, encouraging viewers or participants to see the city as an ongoing ecological collaboration.
- Animism-Materialized: Some media art pieces fuse indigenous ceremonies, living plants, and electronic sensors, enabling the environment itself to "communicate." One might see morphological changes triggered by data from soil or insects, highlighting mutual interdependence.

Through these examples, we see that "Digital Nature" spans a continuum from practical, anthropocentric designs to richly spiritual or philosophical ones—and from purely virtual experiments to physically integrated, "real-world" interventions.

## 5 DISCUSSION: LIMITS, ETHICS, AND FUTURE DIRECTIONS

### 5.1 Ambiguity and the Limits of the Framework

While the above framework clarifies certain patterns, it cannot capture all nuances, especially projects that shift over time or blend contradictory logics. Some large-scale environmental data visualizations may have an underlying socio-political dimension not indicated by a purely morphological or experiential analysis. Furthermore, "Digital Nature" sometimes functions as a buzzword in commercial branding, overshadowing deeper theoretical or ethical concerns. The framework does not fully address potential issues of surveillance or exploitation in "smart" ecological systems, nor does it systematically rank them from ethically sound to ethically dubious.

### 5.2 Supernatural Realities vs. Techno-Occultism

A recurring theme in Digital Nature is a sense of re-enchantment—technologies that appear almost magical. Some scholars embrace this as an opportunity to restore the mystical or sacred dimension stripped away by industrial modernity. Others worry that "techno-occultism" might lead to social mystification, wherein advanced computation is revered as an oracular or omnipotent force. This can obscure the material reality of servers, data labor, or corporate interests behind a veneer of enchanting illusions [36]. Avoiding a regression into uncritical worship of technology requires media literacy, transparency of methods, and socio-cultural critique.

### 5.3 AGI, LLMs, and the Future of Digital Nature

As large language models and multimodal AI systems grow more powerful, we approach scenarios where AI might generate synthetic "natural" environments indistinguishable from real ones—or produce predictive ecological models at scales never before attempted. Hypothetically, an AGI could manage entire ecosystems via sensor networks, adjusting conditions in near real-time. While some see potential for addressing climate challenges, others raise fears of data-driven "nature management" that excludes local or indigenous knowledge. Or, even more radically, AGI might create forms of "nature" beyond conventional physical constraints, ushering in new aesthetic and conceptual frontiers. Yet this begs the question: does transcending or simulating nature at a purely computational level detach us from the tangible, living planet on which we depend?

### 5.4 Political Economy, Decolonization, and Ethics

Critical voices, especially from feminist or indigenous scholars, have noted how digital expansions of "nature" can inadvertently replicate colonial power dynamics [20,34]. If advanced AI systems or AR/VR experiences are developed primarily by powerful corporations or tech-savvy elites, indigenous perspectives risk becoming aesthetic elements without genuine empowerment. Additionally, digital data collection on natural sites can lead to monitoring or privatization that benefits investors more than local communities. Thus, "Digital Nature" cannot be evaluated in isolation from political economy, property rights, and the potential co-optation of spirituality or ecological ethos for profitable spectacle. Genuine decolonization requires that local voices, rituals, and environmental stewardship practices be integrated into every stage of conceptualizing, designing, and governing digital ecologies.

### 5.5 Philosophical Implications: Does Technology Transcend Nature or Coexist with It?

Philosophers since the Enlightenment, from Descartes to Newton, advanced a mechanistic view of nature as an object of control. Later thinkers—such as Prigogine on irreversibility and chaos theory, or Serres on systemic breaks—acknowledged complexity and discontinuity, moving away from absolute determinism. "Digital Nature" works sometimes resurrect the notion that nature can be fully captured or improved by computational means. On the other hand, they may highlight emergent complexities surpassing any single vantage point. Ultimately, this tension is inherent in digital simulations: does modeling and generating nature constitute mastery or an invitation to deeper humility?

Indeed, the increasing hype around "supernatural" technologies or transhuman expansions can perpetuate illusions of total dominion, reminiscent of a modernist dream of conquering nature. Alternatively, well-designed "Digital Nature" projects could foster what some call "co-becoming": an integrated interplay among computational processes, living organisms, and human insight. The direction we take hinges on ethical frameworks, funding priorities, cultural sensitivity, and the willingness to question anthropocentrism.

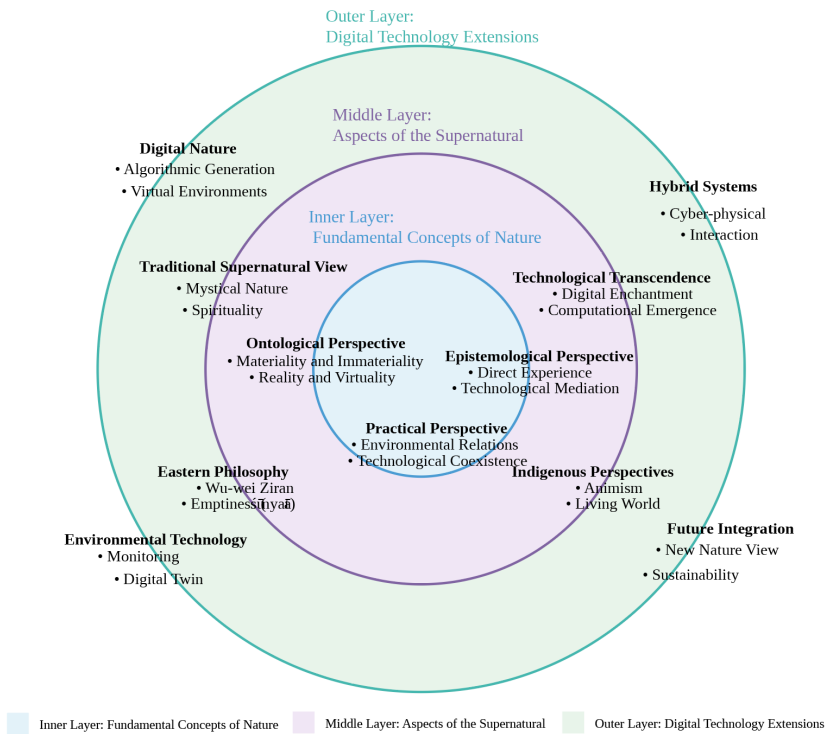

Figure5. Nature and the Supernatural: Theoretical Perspectives

## 6  CONCLUSION

This paper reappraises the sprawling idea of "Digital Nature" by tracing its historical roots, surveying its contemporary uses, and situating it within multiple philosophical traditions. Beginning in the 1980s–1990s with fractals and Perlin Noise, the notion that nature could be algorithmically understood has grown into a wide-ranging conversation involving generative art, bio-art, AI, VR, posthumanism, and spiritual or animistic philosophies. Through a multi-axis framework (Anthropocentric vs. Object-Oriented, Enchanted vs. Implemented, and Simulated vs. Materialized), we have mapped out how diverse projects and research agendas cluster around different priorities—whether they center on human transcendence, industrial optimization, symbiotic multi-agency, or animistic co-creation. This approach underscores that "Digital Nature" is not one singular phenomenon but rather a constellation of overlapping practices and ideas.